\documentclass[%
 reprint,
 amsmath,amssymb,
 aps,
pre,
superscriptaddress,
longbibliography
]{revtex4-2}

\usepackage{graphicx}
\usepackage{dcolumn}
\usepackage{bm}
\usepackage{amsmath}
\usepackage{comment}
\usepackage{bbold}
\usepackage{xcolor}
\usepackage{soul}

\newcommand{\id}{\mathbb{1}} 
\newcommand{\Hcal}{\ensuremath{{\cal{H}}}}

\begin{document}

\preprint{APS/123-QED}

\title{
Diffuse field cross-correlations: scattering theory and electromagnetic experiments
}

\author{Matthieu Davy}
\affiliation{Univ Rennes, INSA Rennes, CNRS, IETR - UMR 6164, F-35000, Rennes, France}

\author{Philippe Besnier}
\affiliation{Univ Rennes, INSA Rennes, CNRS, IETR - UMR 6164, F-35000, Rennes, France}

\author{Philipp del Hougne}
\affiliation{Univ Rennes, INSA Rennes, CNRS, IETR - UMR 6164, F-35000, Rennes, France}
\affiliation{Institut de Physique de Nice, CNRS UMR 7010, Universit\'{e} C\^ote d'Azur, 06108 Nice, France}

\author{Julien de Rosny}
\affiliation{ESPCI Paris, PSL Research University, Institut Langevin, F-75005 Paris, France}%

\author{Elodie Richalot}
\affiliation{ESYCOM lab, Univ Gustave Eiffel, CNRS, F-77454 Marne-la-Vall{\'e}e, France}%

\author{François Sarrazin}
\affiliation{ESYCOM lab, Univ Gustave Eiffel, CNRS, F-77454 Marne-la-Vall{\'e}e, France}%

\author{Dmitry V. Savin}
\affiliation{Department of Mathematics, Brunel University London, Uxbridge, UB8 3PH, United Kingdom}%

\author{Fabrice Mortessagne}
\affiliation{Institut de Physique de Nice, CNRS UMR 7010, Universit\'{e} C\^ote d'Azur, 06108 Nice, France}

\author{Ulrich Kuhl}
\affiliation{Institut de Physique de Nice, CNRS UMR 7010, Universit\'{e} C\^ote d'Azur, 06108 Nice, France}

\author{Olivier Legrand}
\affiliation{Institut de Physique de Nice, CNRS UMR 7010, Universit\'{e} C\^ote d'Azur, 06108 Nice, France}

\date{\today}

\begin{abstract}
The passive estimation of impulse responses from ambient noise correlations arouses increasing interest in seismology, acoustics, optics and electromagnetism. Assuming the equipartition of the noise field, the cross-correlation function measured with non-invasive receiving probes converges towards the difference of the causal and anti-causal Green's functions. 
Here, we consider the case when the receiving field probes are antennas which are well coupled to a complex medium -- a scenario of practical relevance in electromagnetism. We propose a general approach based on the scattering matrix formalism to explore the convergence of the cross-correlation function.  
The analytically derived theoretical results for chaotic systems are confirmed in microwave measurements within a mode-stirred reverberation chamber. This study provides new fundamental insights into the Green's function retrieval technique and paves the way for a new technique to characterize electromagnetic antennas.

\end{abstract}

\maketitle


\section{Introduction}

Even though the field in an ideal disordered medium is diffuse and statistically random, amplitude and phase information about the propagation between two points can be retrieved from the cross-correlation of the field probed at two positions \cite{Berkovits1994,Sebbah2000,Davy2016,rosny2017}. The Green's function retrieval (GFR) technique aims at reconstructing the impulse response between two passive receivers from the cross-correlation of broadband signals. Assuming that the field is at thermal equilibrium, the cross-correlation indeed converges towards the imaginary part of the Green's function between the two field probes \cite{Levin,Rytov,Weaver2001}. This property has been generalized to uniformly distributed noise sources \cite{Snieder2004,Wapenaar2004,Wapenaar2006} with an elegant analogy to time reversal \cite{Tiggelen2003,Derode2003}. GFR is nowadays routinely applied to ambient noise measurements in seismology for surface wave tomography and landslide monitoring~\cite{Shapiro2004,Weaver2005,Campillo2014,LeBreton2021}. The GFR approach also opens the door to passive imaging with sound~\cite{Roux2004,Larose2006b}, microwaves~\cite{Davy2013,Hougne2021_correl} and light~\cite{Badon2015}. In disordered media, including chaotic cavities, multiple scattering of waves enhances the convergence of the cross-correlation since the field illuminating the two probes is ideally a superposition of random plane waves with statistically homogeneous coefficients~\cite{Weaver2001,Tiggelen2003,Larose2006a,Rosny2014}. Because GFR is a self-averaging technique, a single or a few sources are therefore sufficient to recover the ballistic components of the broadband impulse response between two receivers~\cite{Weaver2002,Derode2003,Davy2016,rosny2017,Hougne2021_correl}. 


Reverberation chambers (RC) have become an alternative solution to anechoic rooms to characterize electromagnetic devices~\cite{Besnier2013,Yousaf2020}. Instead of mimicking free-space propagation, the field in mode-stirred RCs is confined within a metallic enclosure which results in multiple reflections off the enclosure's boundaries. Different system realizations are obtained by altering the enclosure's boundary conditions via mechanical~\cite{Madsen2004,Lemoine2008} or electronic~\cite{Kingler2005,sleasman2016microwave,del2018precise,Gros2020} stirring, or via source stirring~\cite{cerri2005source,del2018dynamic}. The field generated by a single antenna (or an antenna array) is typically assumed to be statistically isotropic, uniform and depolarized \cite{Warne2003,Holloway2006} with universal statistics that can be inferred from random matrix theory (RMT) applied to open chaotic systems \cite{Selemani2014,Gros2016,fyod05}. In practice, unstirred-field components may persist but can also be accounted for in the RMT framework~\cite{hart2009effect,yeh2010universal,yeh2010experimental,cozza2011skeptic,savi17,savi20,del2020implementing}. Statistical approaches based on the energy conservation principle enable the extraction of the radiation efficiency or the directivity pattern of transmitting or receiving antennas~\cite{holloway2012reverberation}. In this context, GFR is a technique of potentially great interest to extract the coupling coefficient between two purely receiving antennas~\cite{Davy2016,Hougne2021_correl}. This approach could also accelerate measurements of the coupling coefficients between antennas of a large array since the technique avoids the requirement of turning each antenna successively into its transmitting and receiving modes. Estimating the mutual coupling matrix is key to predicting the channel capacity for communication systems~\cite{Kildal2004,Li2004} and estimating directions of arrival for imaging systems~\cite{Liu2012}. 

In contrast to GFR in acoustics, seismology or optics, the field probes in electomagnetism are typically invasive devices (antennas) that can be well matched to their environment. A detailed analysis of the convergence and the statistics of the cross-correlation function under these conditions (specific to applications in electromagnetism) is to date missing but constitutes an important issue for mode-stirred RCs. In this article, we provide a theoretical analysis of the cross-correlation function in chaotic cavities using the scattering matrix formalism. We show that the cross-correlation function converges, with respect to the stirring process, towards the real part of the transmission coefficient between the two receivers weighted by their reflection coefficients. Our model includes both the coupling of the antennas and losses within the chaotic cavity. The analytically derived average and variance of the cross-correlation are in excellent agreement with measurements taken inside an electromagnetic reverberation chamber using either an array of transmitting sources or a single source and mechanical stirring.

\section{Theory}
\subsection{Ideal cross-correlation function}
\begin{figure}
    \centering
    \includegraphics{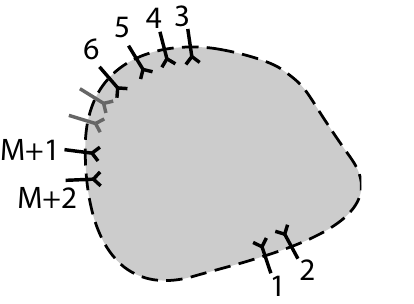}
    \caption{Schematic view of the 2 probes under test and the $M$ emitting sources inside a chaotic cavity.}
    \label{fig:schema}
\end{figure}

A schematic view of our system is shown in Fig.~\ref{fig:schema}. Two receiving probes (indexed $1$ and $2$) and $M$ emitting sources (indexed 3 to $M+2$) are embedded within a chaotic cavity (the RC). The field cross-correlation between probes 1 and 2 estimated from the successive emissions of the $M$ sources is given by
\begin{equation}
C_{12}(\omega) \equiv \Sigma_{m=3}^{M+2} s_{m1}^*(\omega)s_{m2}(\omega),
\label{eq:cross-correl}
\end{equation}
where $s_{mi}(\omega)$ is the complex scattering matrix element (transmission coefficient) between the $m$-th source and the $i$-th probe.

In absence of absorption, the cross-correlation is simply derived from the  flux conservation. Indeed, in such a case, the scattering matrix is unitary, i.e., $S^\dagger(\omega)S(\omega) = \mathbb{1}$. This yields $s_{11}^* s_{12} + s^*_{21} s_{22} +\Sigma_{m=3}^{M+2} s_{m1}^* s_{m2} = 0$, so that $C_{12}(\omega)$ is equal to $C_{0}(\omega)$ defined by
\begin{equation}
C_{12}(\omega)=-[s_{11}^*(\omega) s_{12}(\omega) + s_{22}(\omega) s_{21}^*(\omega)] \equiv C_0(\omega).
\label{eq:convergence}
\end{equation}
%
For pointlike non-invasive probes, such as a sismometer in seismology or a wire much shorter than the wavelength in the microwave domain, Eq.~(\ref{eq:convergence}) is equivalent to the well-known relation between the correlation function and the imaginary part of the Green's function \cite{Davy2016}. Indeed, the magnitude of the reflection coefficient of such probes is strong, $|s_{11}|\sim |s_{22}|\sim1$, such that Eq.~(\ref{eq:convergence}) simplifies to $C_{12} \sim -[s_{12}+s_{21}^*]$ which for systems with time-reversal symmetry indeed corresponds to the real part of the transmission coefficient between the antennas. In the time domain, the inverse Fourier transform $\Tilde{C}_{12}(t)$ of $C_{12}(\omega)$ is then the  superposition of the causal and anti-causal impulse responses, $\Tilde{s}_{12}(t)$ and $\Tilde{s}_{21}(-t)$, respectively, giving a signal which is time symmetric.

Equation~(\ref{eq:convergence}) shows, however, that this symmetry is broken for antennas with different reflection coefficients, $|s_{11}| \neq |s_{22}|$, which occurs when the antennas are not equally coupled to the chaotic cavity. For instance, in the case of a weakly coupled probe ($|s_{22}|\sim1$) and a critically coupled probe ($|s_{11}|\sim 0$), the cross-correlation vanishes at positives times and only the anti-causal response between probes 1 and 2 contributes to the correlation ~\cite{Davy2016}. In the following, we show that in complex media with absorption the function $C_0(\omega)$ still provides a measure of the average cross-correlation function, but with a proportionality factor smaller than unity.

\subsection{Convergence of the average correlation in lossy cavities}

Absorption within an RC is inevitable and breaks the unitary of $S(\omega)$. Hence, in practice, Eq.~(\ref{eq:convergence}) cannot be used to describe the cross-correlation function directly. Helpful insights are provided by a simple heuristic approach assuming that the losses behave as $M_\mathrm{L}$ additional equivalent fictitious channels coupled to the chaotic cavity. Including these $M_\mathrm{L}$ channels within a ``virtual'' scattering matrix $\Hat{S}(\omega)$ of dimension $(M+2+M_\mathrm{L}) \times (M+2+M_\mathrm{L})$ would re-establish its unitarity $\Hat{S}^\dagger(\omega) \Hat{S}(\omega) = \mathbb{1}$. In chaotic cavities, the contributions of the $M$ source antennas and the $M_\mathrm{L}$ fictitious channels to $C_{12}(\omega)$ are assumed to be statistically uncorrelated for different realizations of the diffuse field. In practice, access to different realizations can be implemented via (i) source stirring (altering the position~\cite{cerri2005source} or radiation pattern~\cite{del2018dynamic} of the source antenna(s)), and/or (ii) mode stirring (modifying the boundary conditions of the RC mechanically~\cite{Madsen2004,Lemoine2008} or electronically~\cite{Kingler2005,sleasman2016microwave,del2018precise,Gros2020}). {Averaging over these realizations is denoted by $\langle \ldots \rangle$ in the following.}

As a consequence of the unitarity of $\Hat{S}$ and assuming furthermore that the coupling of the $M_\mathrm{L}$ fictitious channels to the system is the same as that of the $M$ antennas, $\langle s^*_{m1}(\omega)s_{m2}(\omega) \rangle$ becomes a function of $\langle C_0(\omega) \rangle$, $\langle s^*_{m1}(\omega)s_{m2}(\omega) \rangle = \langle C_{0} (\omega) \rangle / (M+M_\mathrm{L})$. We finally obtain 

\begin{equation}
\langle C_{12}(\omega) \rangle = \frac{M}{M+M_\mathrm{L}} \langle C_0(\omega) \rangle 
 = \eta \langle C_0(\omega) \rangle. 
\label{eq:C_0}
\end{equation}

\noindent By construction, the real and positive coefficient $\eta\leq1$ accounts for the unitarity deficit of the scattering matrix caused by the presence of absorption.

In many scenarios in electromagnetic compatibility, antenna characterization and radar imaging, one seeks to characterize the established transmission channel between the receiving antennas, such as line-of-sight or single scattering propagation channels. In these cases,  $\langle C_{0} (\omega) \rangle$ is non-vanishing as the channel can be described as the superposition of the established transmission and a chaotic process within the medium \cite{savi17}. 

We now seek an expression of $\eta$ in terms of parameters that can be easily determined experimentally. We leverage the flux conservation to relate transmissions between the antennas $T_{ij}(\omega) = |S_{ij}(\omega)|^2$ with losses through fictitious channels:
\begin{equation}
1 - \langle T_{11}(\omega) \rangle - \langle T_ {21}(\omega) \rangle = (M+M_\mathrm{L})\langle T_{m1}(\omega) \rangle.
\label{eq:power}
\end{equation}
This equation is a direct consequence of the unitary of $\Hat{S}$. The parameter $\langle T_c(\omega) \rangle = 1 - \langle T_{11}(\omega) \rangle - \langle T_{21}(\omega) \rangle$ characterizes the coupling between antenna $1$ and the chaotic cavity. In particular, $T_c(\omega) = 0$ corresponds to the case of two receiving antennas that are completely isolated. Equation~(\ref{eq:power}) finally leads to
\begin{equation}
\eta = \frac{M\langle T_{m1} \rangle}{1 - \langle T_{11} \rangle - \langle T_{21} \rangle}.
\label{eq:eta}
\end{equation}
Together with $\langle C_{12}(\omega) \rangle = \eta \langle C_0(\omega) \rangle$ (Eq.~(\ref{eq:C_0})), Eq.~(\ref{eq:eta}) provides a full expression of the convergence of the cross-correlation in lossy chaotic cavities.
\subsection{Effective Hamiltonian approach}
In this section, we rigorously confirm the intuitive predictions from the previous section by making use of an effective Hamiltonian formalism to statistically describe the propagation of waves within a chaotic cavity. This formalism proved to be an adequate tool for studying both dynamical and statistical aspects of scattering processes involving resonances \cite{verb85,soko89,fyod97}. The method describes the open quantum or wave system in terms of a non-Hermitian Hamiltonian $\Hcal = H-(i/2)VV^\dagger$, where the Hermitian part $H$ of size $N$ stands for the Hamiltonian of the closed counterpart. The (rectangular) $N \times (M+2)$ matrix $V$ accounts for its coupling to the open channels, which converts the $N$ levels (given by the eigenvalues of $H$) into complex resonances. The scattering matrix is then expressed in a unitary form as
\begin{equation}
    S(E) = \id-i V^\dagger \frac{1}{E - \Hcal} V,
\label{eq:Smatrix}
\end{equation}
where $E$ denotes the scattering energy (or frequency). The resonances correspond to the poles of the S-matrix, which are therefore given by the eigenvalues of $\Hcal$. With this model, the uniform absorption can be easily included, amounting to an imaginary shift of the scattering energy, $E \rightarrow E_\gamma = E + i\Gamma_\mathrm{abs}/2$, where $\Gamma_\mathrm{abs}$ is the absorption width. As a result, the scattering matrix $S(E_{\gamma})$ becomes subunitary and so its unitarity deficit can be quantified by the following exact relation \cite{Savin2003}
\begin{equation}
S^\dagger(E_{\gamma}) S(E_{\gamma}) = \id - \Gamma_\mathrm{abs} Q(E_{\gamma}),
\label{eq:R_operator}
\end{equation}
where $Q$ is the positive-definite Hermitian matrix 
\begin{equation}
Q(E_\gamma) = V^\dagger \frac{1}{(E_\gamma - \Hcal)^\dagger} \frac{1}{E_\gamma - \Hcal}  V.
\label{eq:Q_operator}
\end{equation}
As shown in Ref.~\cite{Savin2003}, Eq.~(\ref{eq:Q_operator}) provides a suitable generalisation of the Wigner-Smith time-delay matrix \cite{Smith1960} in the case of finite absorption. It is worth noting that a factorised structure of Eq.~(\ref{eq:Q_operator}) admits a natural interpretation for a matrix element $Q_{ij}=\psi_i^{\dagger}\psi_j$ as the overlap between the internal parts $\psi_i$ and $\psi_j$ of the wave fields initiated through the channels $i$ and $j$, respectively~\cite{Smith1960,soko97}.

When combined with RMT~\cite{Stoeckmann,guhr98}, the effective Hamiltonian approach provides a powerful tool to describe universal fluctuations in chaotic scattering (see Refs.~\cite{mitc10,fyod11ox} for recent reviews). In particular, exact analytic expressions for a general two-point correlation function of scattering matrix elements were obtained for an arbitrary degree of absorption \cite{fyod05,savi06a} (see also Refs.~\cite{diet10,kuma13,kuma17} for related studies). However, this result cannot be applied to our present problem because it was derived assuming the absence of direct processes (which is a common assumption in RMT). Hence, the term corresponding to Eq.~(\ref{eq:cross-correl}) would simply yield a vanishing contribution. A modification of the conventional RMT formalism is therefore required to incorporate direct processes. Various efforts to capture system-specific features have been reported in the RMT literature. One strategy suggests to incorporate system-specific features via semiclassical calculations of the average impedance matrix in terms of ray trajectories between ports~\cite{hart2009effect,yeh2010universal,yeh2010experimental}. Another approach assumes the presence of an established direct transmission mediated by a resonant mode coupled to the (isolated) complex
environment~\cite{savi17,savi20}. Recently, a hybrid RMT scheme has been proposed to show that one can learn a system-specific pair of coupling matrix and unstirred contribution to the Hamiltonian~\cite{del2020implementing}.

Here, we proceed in the spirit of Refs.~\cite{savi17,savi20} and begin by distinguishing between coupling to the two receiving antennas, $A\equiv\left[V_1 V_2\right]$, and to the $M$ ``bulk'' sources, $B\equiv\left[V_3 ... V_{M+2}\right]$. Moreover, we restrict our study to systems with preserved time-reversal invariance, in line with our experimental setup (see subsequent section for details). Thus, without loss of generality, the $N{\times}(2+M)$ coupling matrix $V = [A\, B]$ is real and $H$ is a real symmetric matrix of size $N \gg 1$. The cross-correlation function $C_{12}(E)$ is then given by
\begin{equation}
    C_{12}(E_\gamma) = A_1^T \frac{1}{(E_\gamma-\Hcal)^\dagger} BB^T \frac{1}{E_\gamma-\Hcal} A_2,
\label{eq:C12}
\end{equation}
where now $\Hcal = H-(i/2)AA^T-(i/2)BB^T$. Note that such a representation can also be used to incorporate non-homogeneous losses in the model~\cite{savi06b}.

Next, 
we assume that the statistical averaging is performed on the sources $B$, so that the system Hamiltonian and the antenna couplings $A_{1,2}$ remain unchanged. This averaging may also include additional stirring which can be effectively used (as discussed in more detail in the experimental section below) to emulate additional sources. Thus, the source number is typically large, $M \gg 1$, but still $M \ll N$. In such a limit, it is natural to consider the $N$-vectors $B_m$ ($m=1,\ldots,M$) as uncorrelated Gaussian random variables with zero mean and second moment $\langle B_{m\mu} B_{m'\mu'}\rangle = \Gamma_m\delta_{mm'}\delta_{\mu\mu'}$. Exploiting their properties~\cite{soko89,soko97}, we can express the averaged function $\langle C_{12}(E_\gamma)\rangle$ in the following form (omitting the higher order terms in $1/M \ll 1$):
\begin{equation}
  \langle C_{12}(E_\gamma) \rangle = \Gamma_\mathrm{src}  
  \left\langle A_1^T \frac{1}{(E_\gamma-\Hcal)^\dagger}\frac{1}{E_\gamma-\Hcal} A_2\right\rangle,
\label{eq:average_C12}
\end{equation}
where $\Gamma_\mathrm{src}=\sum_{m=1}^{M}\Gamma_m$ is the total decay width associated with the channels coupled to the system. This is equivalent to first writing the cross-correlation function as $ C_{12}(E_\gamma) = \psi_1^{\dagger} BB^T \psi_2$ and then assuming that the internal wave fields $\psi_{1,2}$ initiated by the antennas 1 and 2 become (in the large $M$ limit) statistically uncorrelated with the coupling vectors of the source channels,  $\langle \psi_1^{\dagger} BB^T \psi_2 \rangle = \langle \text{Tr}[BB^T] \rangle \langle \psi_1^{\dagger} \psi_2 \rangle$.

The averaged term on the right hand side of Eq.~(\ref{eq:average_C12}) can be readily identified as the off-diagonal element of the averaged time-delay matrix (Eq.~\ref{eq:Q_operator}), resulting in the following elegant relation:
\begin{equation}
    \langle C_{12}(E_\gamma) \rangle = \Gamma_\mathrm{src} \langle Q_{12}(E_\gamma)\rangle.
\label{eq:C_Q}
\end{equation}
In order to further establish a connection with the function $C_0(E_\gamma)$ that can be cast as $C_0(E_\gamma)=-\psi_1^{\dagger} AA^T \psi_2$, we take the off-diagonal element of relation~(\ref{eq:R_operator}) and find
\begin{equation}
C_{12}(E_\gamma) = C_0(E_\gamma) - \Gamma_\mathrm{abs} Q_{12}(E_\gamma).
\label{eq:C12_unitarity}
\end{equation}
Finally, averaging this equation over different realizations and making use of Eq.~(\ref{eq:C_Q}), we obtain the representation of the average cross-correlation function with absorption as
\begin{equation}
\langle C_{12}(E_\gamma) \rangle= \frac{\Gamma_\mathrm{src}}{\Gamma_\mathrm{abs} + \Gamma_\mathrm{src}} \langle C_0(E_\gamma)\rangle = \eta \langle C_0(E_\gamma)\rangle .
\label{eq:C12_unitarity_av}
\end{equation}
This equation fully confirms the result of Eq.~(\ref{eq:C_0}). For the sake of comparison with the heuristic argument of the previous section, one can identify $M \langle T_{m1}\rangle = \Gamma_{\text{src}}$ and $M_\mathrm{L} \langle T_{m1}\rangle = \Gamma_{\text{abs}}$ (both widths being expressed here in units of the mean density of states). 

Our approach also allows us to further relate the above result to Eqs.~(\ref{eq:power}) and (\ref{eq:eta}). The first diagonal element of $S^\dagger(E_{\gamma}) S(E_{\gamma})$ given in Eq.~(\ref{eq:R_operator}) yields
\begin{equation}
    T_{11}(E_\gamma) + T_{12}(E_\gamma)  +\sum_{m=3}^{M+2}  T_{m1}(E_\gamma)  =  1  - \Gamma_\mathrm{abs} Q_{11}(E_\gamma).
\label{eq:C_Q1}
\end{equation}
\noindent The same arguments used to derive Eq.~(\ref{eq:average_C12}) then provide  $\Gamma_\mathrm{src} \langle Q_{11}(E_\gamma) \rangle = \sum_{m=3}^{M+2} \langle T_{m1}(E_\gamma) \rangle$ so that we finally obtain
\begin{equation}
    \sum_{m=3}^{M+2} \langle T_{m1}(E_\gamma) \rangle = \eta [ 1 - \langle T_{11}(E_\gamma) \rangle -\langle T_{12}(E_\gamma) \rangle ],
\end{equation}
which is equivalent to Eq.~(\ref{eq:eta}).

The approach based on Eq.~(\ref{eq:C12}) provides a route to study cross-correlations in a systematic and controllable way, which may include, for instance, higher order corrections or possible fluctuations of absorption rates.

\subsection{Fluctuations of the cross-correlation}

Having established the convergence of the average cross-correlation, we now evaluate its fluctuations and the corresponding signal-to-noise ratio (SNR). Simple relations can be found using the heuristic model of $M$ source antennas and $M_\mathrm{L}$ fictitious channels that describe absorption and the assumption that the transmission coefficients $s_{m1}$ and $s_{m2}$ are random variables with Gaussian distribution. The relative variance of the fluctuations, $\langle |\delta C_{12}(\omega)|^2 \rangle = \langle |C_{12}(\omega) - \eta C_{0}(\omega)|^2 \rangle$, is (see Appendix for derivation)

\begin{equation}
\frac{\langle |\delta C_{12}(\omega)|^2 \rangle}{|\langle C_{12}(\omega) \rangle|^2} = \frac{1-\eta}{M} \frac{[1 - \langle T_{11} \rangle - \langle T_{21} \rangle]^2}{|\langle C_{0}(\omega) \rangle|^2}.
\label{eq:variance_eq_end}
\end{equation}

In absence of absorption, the unitary scattering matrix gives $\eta=1$ and $\langle |\delta C_{12}(\omega)|^2 \rangle=0$, as the reconstruction is expected to be perfect (see Eq.~(\ref{eq:convergence})). In the strong absorption limit ($\eta \rightarrow 0$), Eq.~(\ref{eq:variance_eq_end}) shows that the relative fluctuations barely depend on absorption as long as the chaotic nature of the RC is preserved. Moreover, this equation highlights the impact of the coupling between the system of the two antennas and the cavity. For isolated antennas forming an enclosure within the larger chaotic cavity, i.e. an almost isolated $2\times2$ sub-system of receiving antennas, $T_c \rightarrow 0$ and the fluctuations are small. This is, for instance, the case when a line-of-sight propagation between directive antennas gives rise to strong direct coupling between the antennas.

Equations~(\ref{eq:C_0}), (\ref{eq:eta}) and (\ref{eq:variance_eq_end}) provide a framework to analyze the convergence of the correlation function in chaotic systems.

\section{Experimental results}

Having established the theoretical predictions for the convergence and the variance of the cross-correlation function in chaotic cavities, we confirm these results in this section with two experimental setups for which the degree of control $\eta$ varies from $\eta \rightarrow 1$ to $\eta \rightarrow 0$.

\subsection{Two-dimensional cavity}

\begin{figure}
\includegraphics[width=8cm]{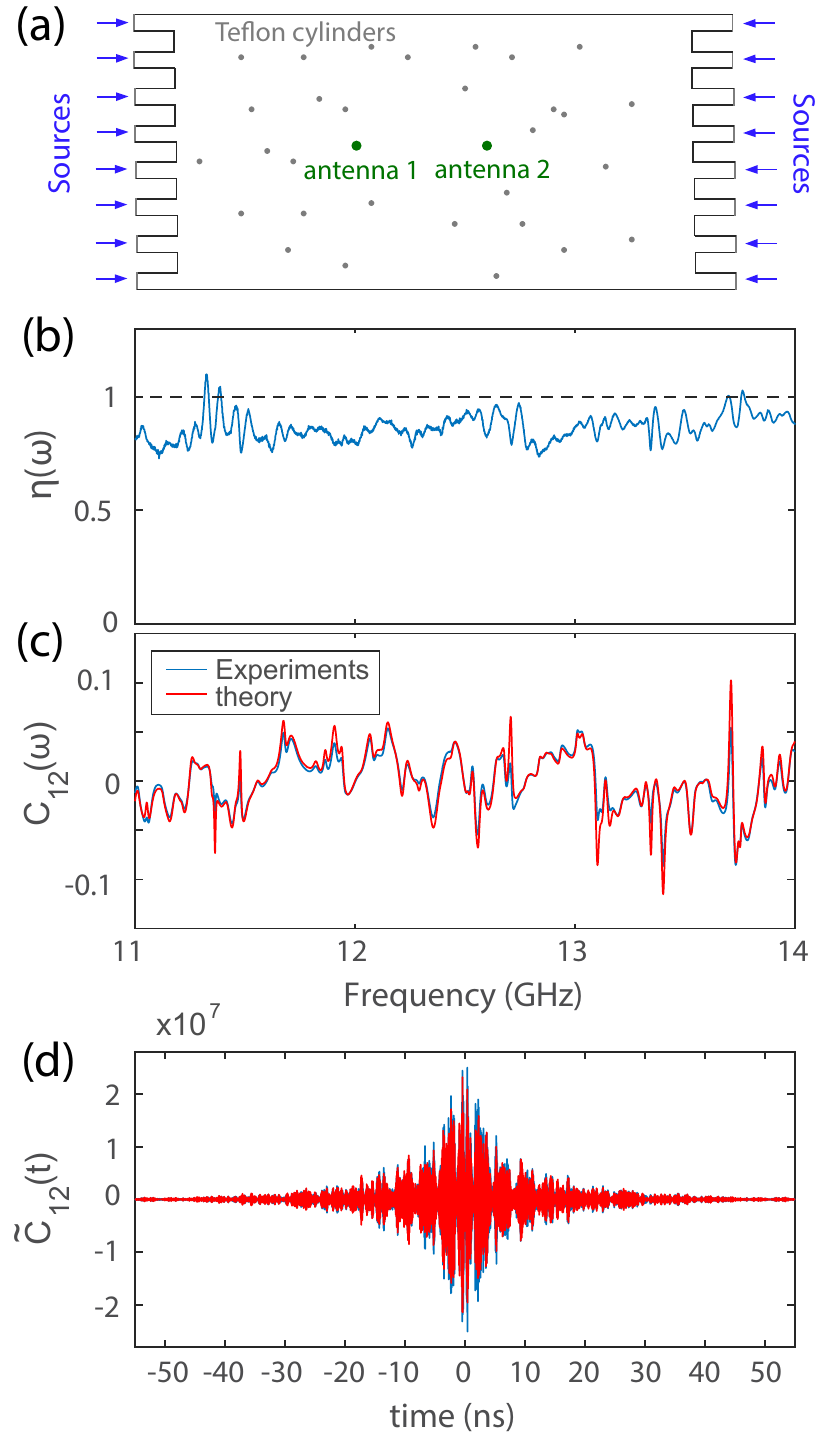}
\caption{\label{fig:setup_2D} (a) Schematic drawing of the 2D experimental setup used to reconstruct the impulse response between two wire antennas located near the middle of the cavity. The spacing between the wire antennas is $d = 0.06$~m. {(b) Variations of $\eta(\omega)$ with frequency.}
(c) Real part of the cross-correlation as a function of frequency, $C_{12}(\omega)$ (blue), and corresponding results from Eqs.~(\ref{eq:C_0}) and (\ref{eq:eta}) (red). (d) Time domain representation of the results displayed in (c).}
\end{figure}

The first setup is an electromagnetic cavity of length $L=0.5$~m, width $L=0.25$~m and height $h = 0.008$~m (see Fig.~\ref{fig:setup_2D}(a) and Ref.~\cite{Davy2021}). The cavity supports a single mode in its vertical polarization so that the setup emulates a two-dimensional cavity. The openings of the cavity are fully controlled with $M=16$ single-channel waveguides (coax-to-waveguide transitions) that are nearly perfectly matched to the cavity in the frequency range of interest 11-14 GHz. The single-channel waveguides act as sources to measure the cross-correlation function between two wire antennas separated by 0.06~m and located near the middle of the cavity as shown in Fig.~\ref{fig:setup_2D}(a). These receivers are coupled to the cavity through holes drilled into the top plate of the cavity. 
Unlike open waveguides, the space between neighbouring single-channel waveguides is metallic such that the amount of reverberation within the system is significantly enhanced.
The wave field within the cavity is additionally randomized using 30 randomly distributed teflon cylinders. Measurements of the transmission coefficients $s_{m1}(\omega)$ and $s_{m2}(\omega)$ between the sources and the wire antennas are carried out with two 8-port electro-mechanical switches connected to a vector network analyzer (VNA).

As the system's decay rate is dominated by leakage through the attached scattering channels, the scattering matrix is almost unitary and the degree of control $\eta(\omega)$ approaches unity. {Figure~\ref{fig:setup_2D}(b) shows that $\eta(\omega)$, computed using Eq.~(\ref{eq:eta}), fluctuates over the frequency range. Note that the parameter $\eta$ is theoretically bounded by unity and values of $\eta(\omega)$ larger than unity found experimentally are due to small calibration and estimation errors.} 
Its average over the frequency range is $\langle \eta \rangle_\omega \sim 0.86$. {In this case, the cross-correlation function directly provides $\eta(\omega) C_0(\omega)$ without the need to average over different realizations as its fluctuations around $\eta(\omega) C_0(\omega)$ are small}. The cross-correlation $C_{12}(\omega) = \Sigma_{m=3}^{M+2} s^*_{m1}(\omega) s_{m2}(\omega)$ shown in Fig.~\ref{fig:setup_2D}(c) is in excellent agreement with the theoretical prediction $\eta(\omega)C_0(\omega)$. The temporal representation of the cross-correlation $\Tilde{C}_{12}(t)$, obtained via an inverse Fourier transform of $C_{12}(\omega)$, is presented in Fig.~\ref{fig:setup_2D}(d) and also confirms that the impulse response between the wire antennas is accurately reconstructed over more than 50~ns. 


\begin{figure}
\includegraphics[width=8cm]{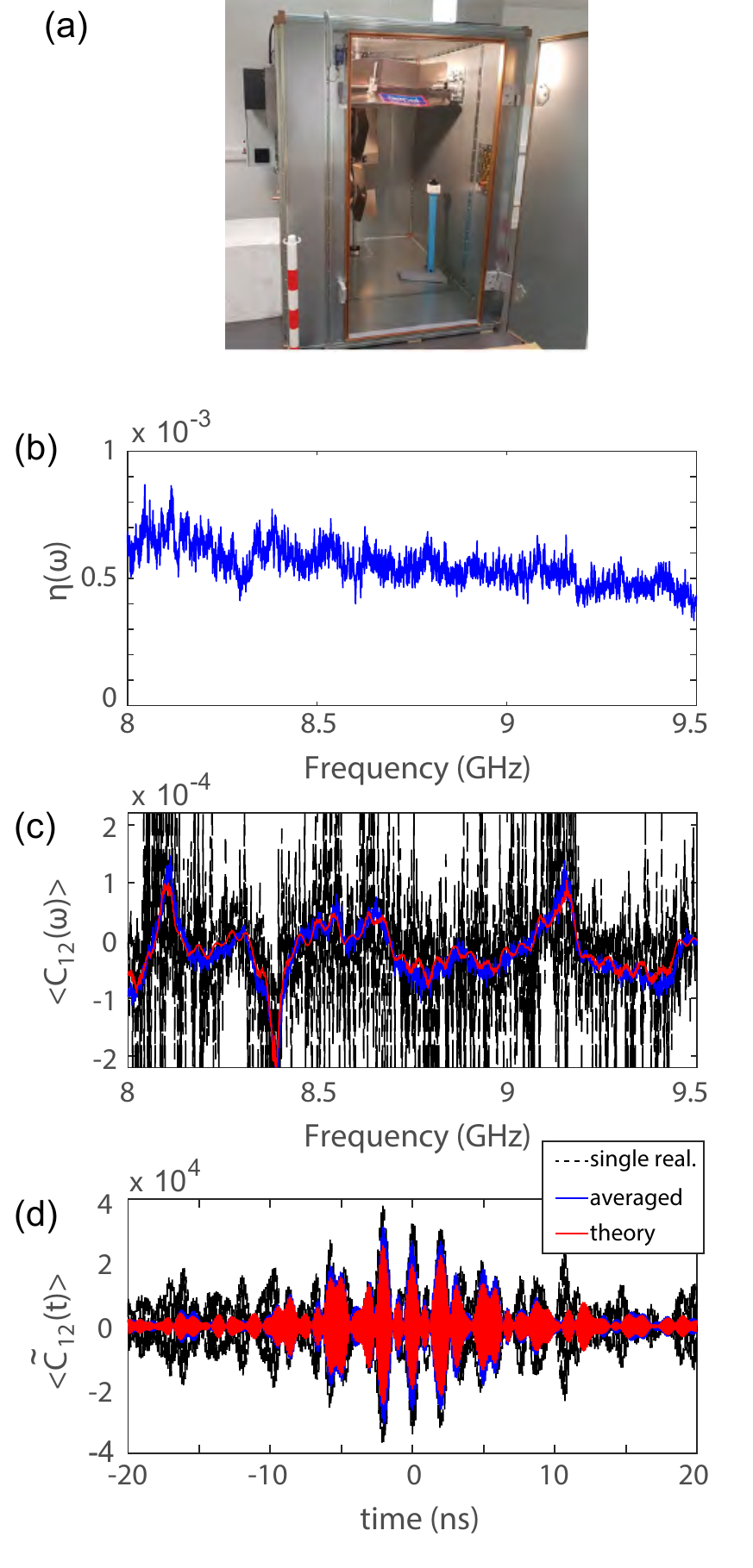}
\caption{\label{fig:3D} (a) Photographic image of the RC. (b) Variations of $\eta(\omega)$ with frequency obtained experimentally for a spacing between the two receiving horn antennas of $d = 0.55$~m.
(c) Corresponding real part of the cross-correlation $Re[C_{12}(\omega)]$  {for a single realization (black dashed line), an averaging over 200 realizations of the mode-stirrer} (blue line) and the result from Eqs.~(\ref{eq:C_0}) and (\ref{eq:eta}) (red line).
(d) Time domain representation of the results displayed in (c).}
\end{figure}

\subsection{Three-dimensional reverberation chamber}

We now consider (in contrast to the previous subsection) the case of a large cavity for which absorption within the system clearly dominates over losses induced by the antennas. This second setup is a three-dimensional RC of volume $5.25~\text{m}^3$, depicted in Fig.~3(a). The two receiving antennas are two horn antennas facing each other and the source is a third horn antenna located near a corner of the RC. Measurements of the transmission coefficients $s_{ij}(\omega)$ are carried out over the $8-9.5$~GHz range with a VNA. Two mechanical stirrers of large dimensions are rotated to realize (and average over) independent realizations. The quality factor of the cavity $Q = 2\pi f_0 \tau$ is found by fitting in the time domain the exponential decay of the transmitted intensity between two antennas, $\langle |\Tilde{s}_{ij}(t)|^2 \rangle\sim e^{-t/\tau}$. We estimate that $\tau \sim 259$~ns and $Q\sim 14250$ around $f_0 = 8.75$~GHz.

Given the large size of the RC, losses through the antennas are small in comparison to absorption within the RC. The parameter $\eta(\omega)$ {shown in Fig.~\ref{fig:3D}(b)} fluctuates between $10^{-4}$ and $10^{-3}$ over the considered frequency range and for random configurations of the RC. The degree of control provided by a single antenna is therefore small as the scattering matrix is far from being unitary. The benchmark function $C_0(\omega)$ is poorly reconstructed from a single correlation $C_{12}(\omega) = s_{31}^*(\omega) s_{32}(\omega)$ {as large fluctuations are observed (see Fig.~\ref{fig:3D}(b))}. However, averaging of $C_{12}(\omega)$ over $N_{\mathrm{r}} = 200$ realizations of the stirrer provides an excellent reconstruction of $\langle C_0(\omega) \rangle$. The inverse Fourier transform $\langle C_{12}(t) \rangle$ is also in very good agreement with $\langle C_{0}(t) \rangle$ for $|t|<15$~ns.

\begin{figure}
\includegraphics[width=8.5cm]{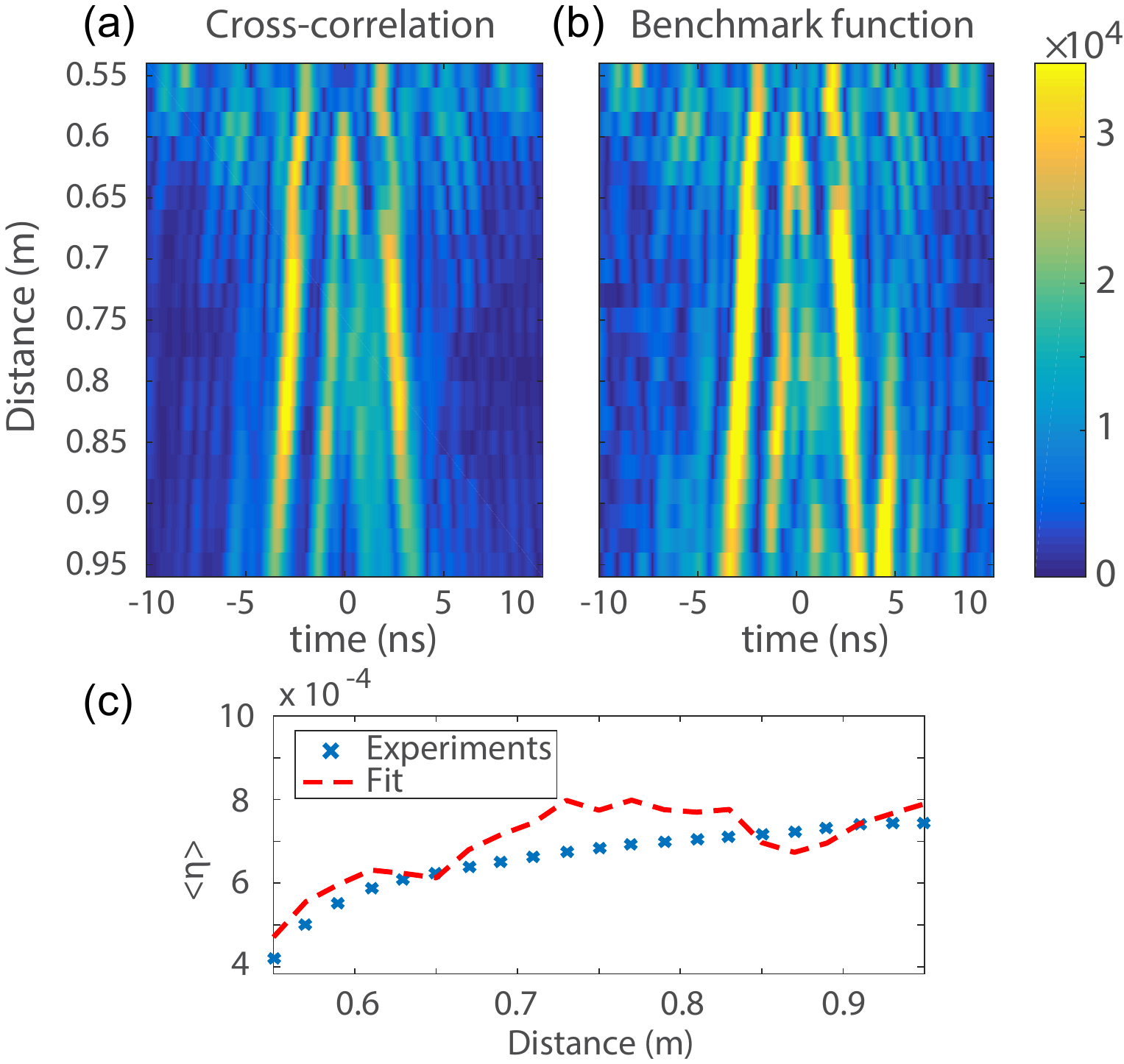}
\caption{\label{fig:bscan} The variations of the envelope of the correlation function $\Tilde{C}_{12}(t)$ (a) and $\eta \Tilde{C}_{0}(t)$ (b) in the time domain are shown with respect to the distance $d$ between the antennas. 
(c) The coefficient $\langle \eta \rangle_\omega$ found using Eq.~(\ref{eq:eta}) is compared to the coefficient giving the best fit of $\langle C_{12}(\omega) \rangle$ with $\eta \langle C_0(\omega) \rangle$.}
\end{figure}

\begin{figure}
\includegraphics[width=8.5cm]{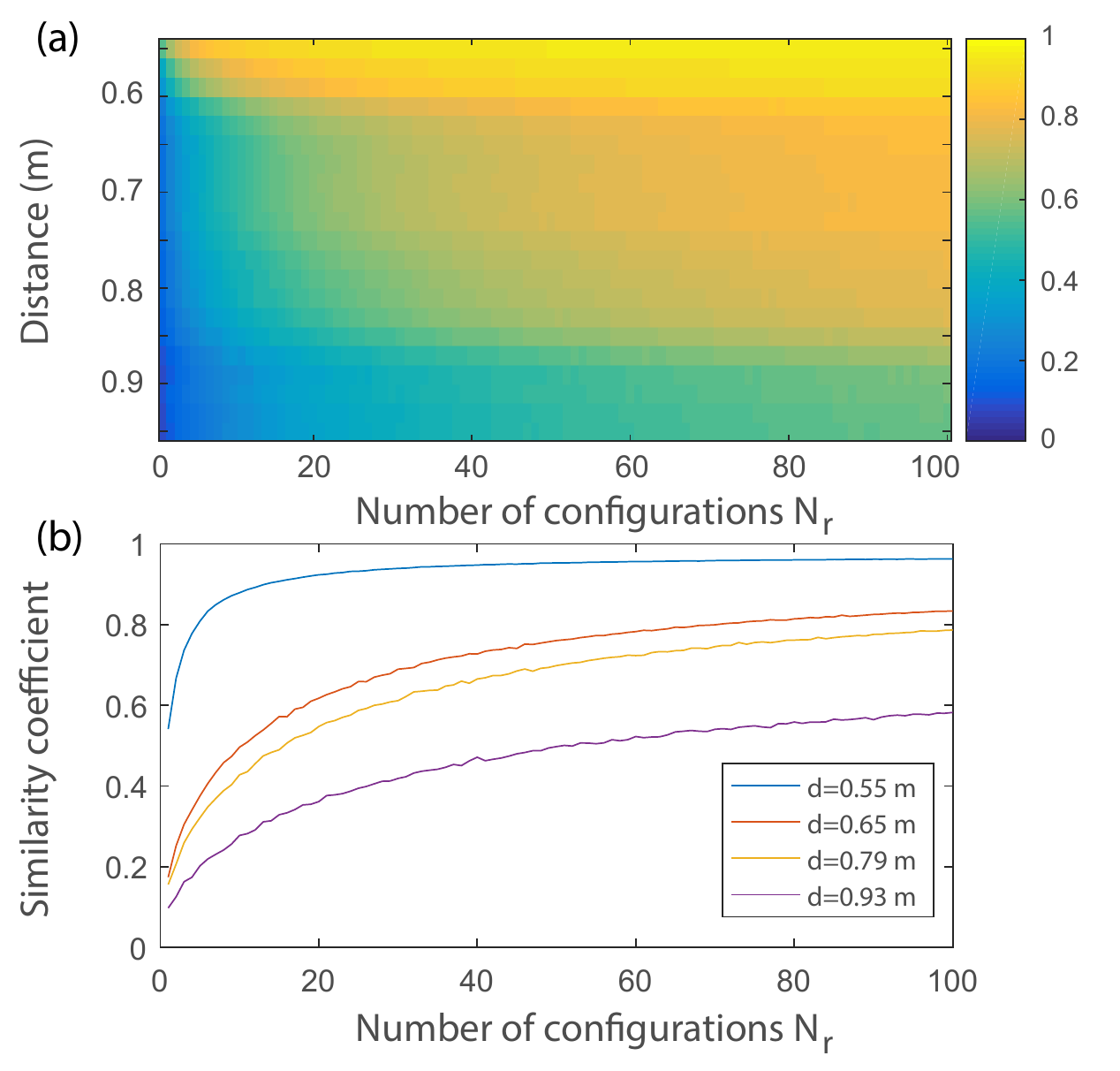}
\caption{\label{fig_similarity}(a) Colormap of the similarity coefficient $S(N_r,d)$ shown as a function of the number of configurations $N_r$ over which the averaging is performed and the distance $d$ between the receiving antennas.}
\end{figure}

We now explore the convergence of the cross-correlation as a function of the distance $d$ between the two receiving horn antennas, and hence on the strength of the established transmission (line-of-sight propagation). We consider $0.55\ \mathrm{m}<d<0.95$~m. The average cross-correlation $\langle C_{12}(t) \rangle$ faithfully reproduces $\langle C_0(t) \rangle$, even though the reconstruction slightly deteriorates as $d$ increases (see Fig.~\ref{fig:bscan}(a,b)). Only the pulses with maximal energy corresponding to ballistic propagation between the antennas are accurately reconstructed for $d=0.95$~m {as the average coupling coefficient between them is small and the cross-correlation is dominated by the noise level}. The coefficient $\langle \eta(\omega) \rangle_\omega$ was defined in Eq.~(\ref{eq:eta}) as the overall amplitude factor relating $\langle C_{12}(\omega) \rangle$ and $\langle C_{0}(\omega) \rangle$. This coefficient is found to be in good agreement with the value found from the best fit of $\langle C_{12}(\omega) \rangle$ to $\langle C_{0}(\omega) \rangle$ (see Fig.~\ref{fig:bscan}(c)). Our experiments therefore fully confirm our theoretical predictions on the convergence of the cross-correlation function from the previous section.

To explore the deterioration of the reconstruction as the distance between antennas increases, we compute the similarity (Pearson's correlation coefficient) between the experimental results and the theoretical prediction $\eta(\omega)C_0(\omega)$:

\begin{equation}
S(N_r)=\frac{\int_{\omega_{\text{min}}}^{\omega_{\text{max}}}  d\omega \langle C_{12}(\omega) \rangle_{N_r} [\eta(\omega) C_0^*(\omega)]}{\sqrt{\int_{\omega_{\text{min}}}^{\omega_{\text{max}}} d\omega |\langle C_{12}(\omega) \rangle_{N_r}|^2} \sqrt{ \int_{\omega_{\text{min}}}^{\omega_{\text{max}}} d\omega |\eta(\omega) C_0^*(\omega)|^2 }} ,
\label{eq:similarity}
\end{equation}

\noindent where $\langle ... \rangle_{N_r}$ denotes an averaging over $N_r$ realizations of the RC. Figure~\ref{fig_similarity} shows that for the smallest considered antenna separation, $S(N_r)$ almost converges towards its maximum $S=0.96$ for $N_r>30$. The reconstruction of the impulse response between two nearby antennas is hence excellent even if only a small number of uncorrelated realizations is used. However, the similarity converges more slowly with respect to $N_r$ as $d$ increases. Equation~(\ref{eq:variance_eq_end}) indeed shows that the fluctuations of $C_{12}(\omega) / \langle C_{12}(\omega) \rangle$ scale as $1/| \langle C_{0}(\omega) \rangle |^2$. When the antennas separation is increased, the direct coupling and $| \langle C_{0}(\omega) \rangle |^2$ decrease so that averaging over more independent realizations is required to achieve the same value of similarity.

\begin{figure}
\includegraphics[width=8.5cm]{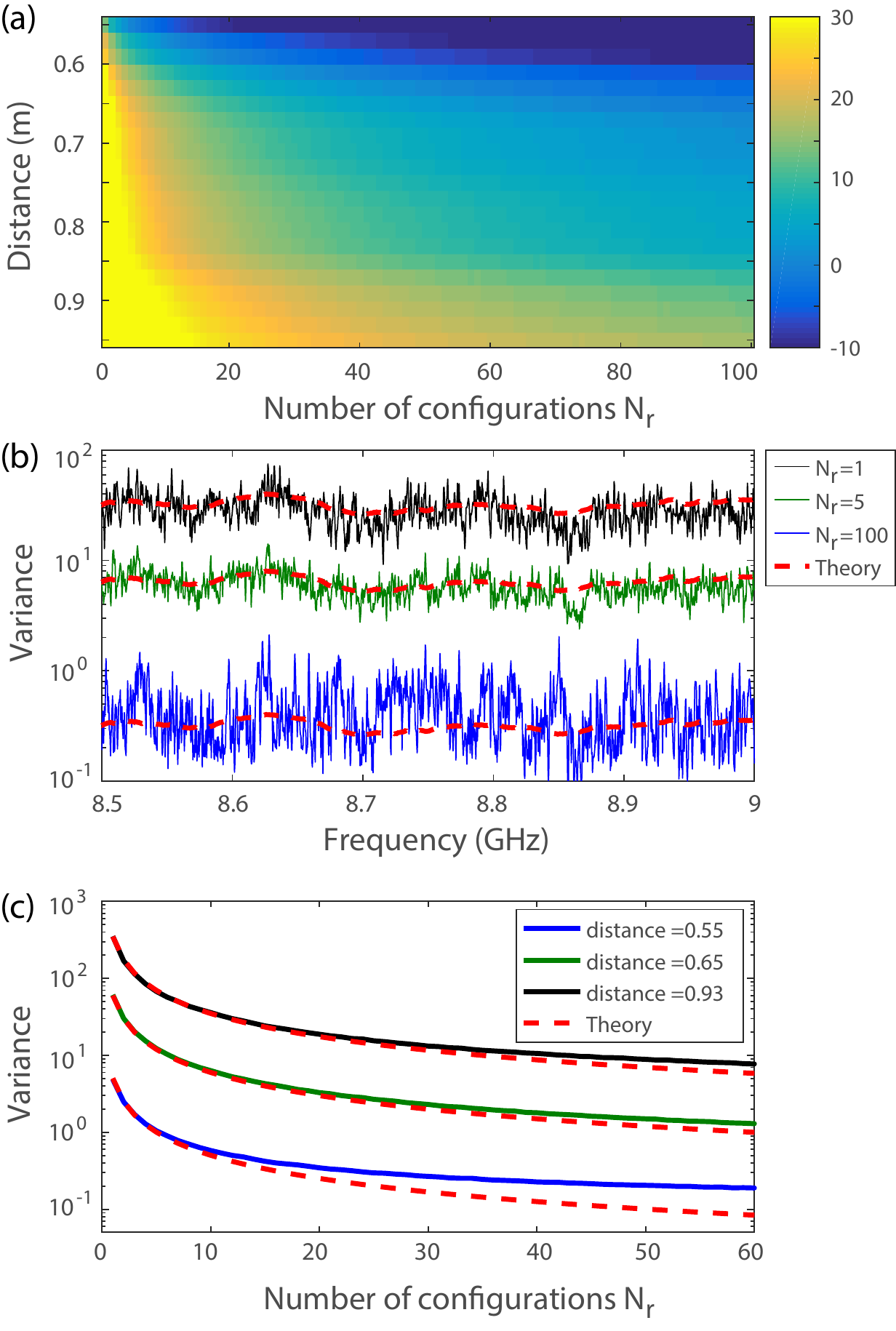}
\caption{\label{fig_variance} (a) Colormap of the variance $V(\omega,d,N_r)$ shown as a function of the number of configurations $N_r$ over which the averaging is performed and the distance $d$ between the receiving antennas. (b) Spectra of the variance, shown between 8 and 9 GHz, for $N_r=1$ (black line), $N_r=5$ (green line) and $N_r=100$ (blue line) are compared to theoretical predictions (red dashed lines). (c) $V(\omega,d,N_r)$ is shown as a function of $N_r$ for three distances between the antennas: $d=0.55$~m (blue line), $d=0.65$~m (green line) and $d=0.93$~m (black line). The red curves are theoretical predictions.}
\end{figure}

Finally, we compute the variance $V(\omega,d,N_r)$ which is estimated for each $N_r$ using 50 random permutations of $N_r$ configurations of the stirrer over a maximum of 200 configurations. The variations of $V(\omega,d,N_r)$ with $d$ and $N_r$ are presented in Fig.~\ref{fig_variance}. As expected, this variance is minimal for small antenna separations. To compare experimental results with theoretical predictions, the number of source antennas $M$ is replaced in Eq.~(\ref{eq:variance_eq_end}) by the number of independent realizations of the cavity $N_r$ (see Fig.~\ref{fig_variance}(b)). These two degrees of freedom provide equivalent independent states of the cavity leading to the convergence of the cross-correlation towards $\eta \langle C_0 \rangle$. We therefore expect that they have the same impact on the convergence of the cross-correlation function. The spectra of $V(\omega,d,N_r)$ for different values of $N_r$ are indeed seen to be in very good agreement with the theory.

The decrease of $\langle V(\omega,d,N_r) \rangle_\omega$ with $N_r$ shown in Fig.~\ref{fig_variance}(c) also closely coincides with the theoretical prediction for $N_r<20$. However, for larger $N_r$, we observe a departure between theoretical and experimental curves, the experimental decay rate being smaller than the theoretical one. 
{This comes as no surprise since the similarity coefficient between the theoretical and experimental results does not reach unity even for the smallest distance, as seen in Fig.~5(b). Residual effects that are not included within our model therefore impact the cross-correlation function. We identify several processes that may degrade the estimation of the benchmark function. Firstly, a slightly incorrect estimation of $\eta(\omega)$ may be due to experimental calibration errors. Secondly, the residual presence of correlated field components for the different realizations of the stirring process reduces the degree of independence between subsequent measurements and can be quantified as follows: $|\langle S_{13}(f) \rangle|^2 / \langle |S_{13}(f)|^2 \rangle = 0.07$. The effective number of independent realizations is therefore smaller than $N_r$. The two aforementioned effects lead to a small additive non-averaging contribution $C_{\text{err}}(\omega)$ to the measured cross-correlation, $C_{12}(\omega) = \eta(\omega) C_0(\omega) + \delta C_{12}(\omega) + C_{\text{err}}(\omega)$. The variance estimated experimentally therefore converges towards  $\langle V(\omega,d,N_r) \rangle_\omega = |C_{\text{err}}(\omega)|^2 / |\eta(\omega) C_0(\omega)|^2 $ for $N_r \rightarrow \infty$ instead of vanishing. The impact of $C_{\text{err}}(\omega)$ is small for small values of $N_r$ but dominates the noise level for large $N_r$, explaining the saturation of the variance in Fig.~\ref{fig_variance}(c).}


\section{Conclusion}

In conclusion, our study provides a framework to analyze the cross-correlation function measured in electromagnetic RCs with well coupled antennas. {Using a formalism based on the scattering matrix, we have shown  that the average cross-correlation function that can be measured between two receiving antennas inside a chaotic cavity illuminated with a single source or an array of sources depends on two parameters. First, the function $C_0(\omega)$ is related to the coupling between the two receiving antennas; second, the parameter $\eta$ is a proportionality factor related to absorption within the cavity which leads to the non-unitarity of the scattering matrix. As the system's decay rate is dominated by absorption instead of leakage through the attached scattering channels, fluctuations of the cross-correlation function however degrade the estimation of the transmission coefficient between the antennas; the variance mainly depends on the coupling between the receiving antennas and the cavity. In this case, averaging over independent realizations of the cavity is needed, and the variance decreases inversely to the number of realizations.} Our derived theoretical results are nicely confirmed by our experimental measurements carried out in 2D and 3D electromagnetic cavities. {Due to the universality of the scattering matrix approach, our results apply not only to chaotic cavities but also to other complex media such as random multiple-scattering samples.}

We expect our study to result in new perspectives and applications in electromagnetic compatibility, the characterization of antennas and electromagnetic objects. {Firstly, the cross-correlation technique provides a way to accurately measure the coupling between two receiving antennas that cannot be turned into their emitting modes. This may be crucial especially in the context of the internet of things for which small embedded sensors are employed. Secondly, the technique in reverberation chambers may accelerate measurement of the mutual coupling matrix for large antenna arrays. A complete measurement indeed requires the field transmission coefficients between each pair of antennas and is highly challenging as each antenna must be successively turned into its transmitting and receiving modes. However, retrieving these transmission coefficients with all antennas in their receiving modes may provide a great simplification of the setup and a speed-up of the characterization process.}

\begin{acknowledgments}
This publication was supported by the French Agence Nationale de la Recherche under reference ANR-17-ASTR-0017, the European Union through the European Regional Development Fund (ERDF), by the French region of Brittany and Rennes M{\'e}tropole through the CPER Project SOPHIE/STIC \& Ondes. M.D. acknowledges the Institut Universitaire de France (IUF).
\end{acknowledgments}

\section{Appendix}

Here, we derive an expression for the fluctuations of the cross-correlation function $\delta C_{12} = C_{12} - \eta C_0$. For the sake of simplicity, the angular frequency $\omega$ is omitted in the following equations. We begin by expressing $\delta C_{12}$ as
\begin{equation}
\delta C_{12} = (1-\eta) \Sigma_{m=3}^{M+2} s^*_{m1}s_{m2} - \eta \Sigma_{m'=1}^{M_\mathrm{L}} s^*_{m'1}s_{m'2},
\end{equation}
where the index $m$ indicates the contribution of a source antenna and the index $m'$ the contribution of a fictitious lossy channel. This leads to
\begin{equation}
\begin{split}
\langle |\delta C_{12}|^2 \rangle & = (1-\eta)^2 \langle |\Sigma_{m=3}^{M+2} s^*_{m1}s_{m2}|^2 \rangle \\ 
& + \eta^2 \langle |\Sigma_{m'=1}^{M_\mathrm{L}} s^*_{m'1}s_{m'2}|^2 \rangle \\
& - \eta (1-\eta)\langle [\Sigma_{m=3}^{M+2} s^*_{m1}s_{m2} \Sigma_{m'=1}^{M_\mathrm{L}} s_{m'1}s_{m'2}^* + c.c]\rangle
\end{split}.
\label{eq:App_decomp_variance}
\end{equation}
Here c.c. means complex-conjugate of the previous term. We first estimate the term $\langle |\Sigma_{m=3}^{M+2} s_{m1}^*s_{m2} |^2 \rangle = M \langle s_{m1}^* s_{m1} s_{m2} s_{m2}^*\rangle + M(M-1) \langle s_{m1}^* s_{m2} s_{m'1} s^*_{m'2}\rangle$. We assume that the coefficients $s_{m1}$ and $s_{m2}$ are Gaussian variables with zero mean and the same variances, which is the case in chaotic cavities with strong absorption~\cite{kuma13}. This gives $\langle s_{m1}^* s_{m1} s_{m2} s_{m2}^*\rangle = \langle T_{m1} \rangle \langle T_{m2} \rangle+ |\langle s_{m1}^* s_{m2} \rangle|^2$. We have previously shown that the transmission from the receivers is given by $\langle T_{m1} \rangle =\langle T_{m2} \rangle = \eta T_c /M $. The second term is related to the cross-correlation function as $M^2 |\langle s_{m1}^* s_{m2} \rangle|^2 = \eta^2 |\langle C_0 \rangle|^2$ so that
\begin{equation}
(1-\eta)^2 \langle |\Sigma_{m=3}^{M+2} s_{m1}^*s_{m2}|^2 \rangle = (1-\eta)^2 \eta^2 [\frac{T_{c}^2}{M} + |\langle C_0 \rangle|^2].
\end{equation}
A similar relation can be established for the sum over the fictitious lossy channels. Using that $M_\mathrm{L} = M(1-\eta)/\eta$, we obtain
\begin{equation}
(1-\eta)^2 \langle |\Sigma_{m'=1}^{M_\mathrm{L}} s_{m'1}^*s_{m'2}|^2 \rangle = (1-\eta)^2 \eta^2 [\eta \frac{T_{c}^2}{M(1-\eta)} + |\langle C_0 \rangle|^2].
\end{equation}
Finally, to evaluate the last term in Eq.~(\ref{eq:App_decomp_variance}), we assume that the average correlation between two channels $\langle s_{m1} s_{m'1} \rangle$ vanishes. This yields $\langle \Sigma_{m=3}^{M+2} s_{m1}^*s_{m2} \Sigma_{m'=1}^{M_\mathrm{L}} s_{m'1}s_{m'2}^* \rangle = M M_\mathrm{L} |\langle s_{m1}^* s_{m2}\rangle|^2$ and
\begin{equation}
\langle \Sigma_{m=3}^{M+2} s_{m1}^*s_{m2} \Sigma_{m'=1}^{M_\mathrm{L}} s_{m'1}s_{m'2}^* \rangle = \eta(1-\eta) |\langle C_0 \rangle|^2.
\end{equation}
Using Eqs.~(\ref{eq:C_0}) and (\ref{eq:eta}), we finally obtain the result presented in Eq.~(\ref{eq:variance_eq_end}).

\bibliographystyle{apsrev4-1}


\providecommand{\noopsort}[1]{}\providecommand{\singleletter}[1]{#1}%

\end{document}